# Polarization-sensitive tunable extraordinary terahertz transmission based on a hybrid metal-vanadium dioxide metasurface

**S. Hadi Badri[1], Sanam SaeidNahaei[2], Jong Su Kim[2]**

[1] *Department of Electrical Engineering, Sarab Branch, Islamic Azad University, Sarab, Iran.*
[2] *Department of Physics, Yeungnam University, Gyeongsan, 38541, Republic of Korea.*
[*] *sanam.nahaei@yu.ac.kr*

**Abstract:** A thermally tunable extraordinary terahertz transmission in a hybrid metal-vanadium dioxide ($VO_2$) metasurface is numerically demonstrated. The metasurface consists of a metal sheet perforated by square loops while the loops are connected with strips of $VO_2$. The frequency and amplitude of the transmission resonance are modulated by controlling the conductivity of the $VO_2$. For *y*-polarized incident field, the resonance transmission peak redshifts from 0.88 to 0.81 THz upon insulator-to-metallic phase transition of $VO_2$. For *x*-polarized incident field, the transmission resonance at 0.81 THz is observed in the insulator phase. However, in the metallic phase of $VO_2$, the electromagnetic field is effectively reflected in the 0.5-1.1 THz range with a transmission level lower than 0.14. The proposed metasurface can be utilized as a terahertz modulator, reconfigurable filter, or switch.

## 1. Introduction

The terahertz regime of the electromagnetic spectrum ranges from 0.1 to 10 THz. In the frequencies below a few hundred GHz, the electromagnetic devices relying on the motion of free electrons are dominant. At the higher frequencies, i.e., infrared to ultraviolet range, controlling the motion of photons forms the basis of electromagnetic devices. In the terahertz gap, between these two regions, the efficiency of the electronic and photonic devices is low. Metamaterials, as artificial structures with properties not found in natural materials, play a crucial role in designing terahertz devices [1, 2]. Metamaterials and metasurfaces, the two-dimensional counterparts of metamaterials, have been employed to develop unique applications such as filters [3, 4], cloaking [5, 6], lenses [7, 8], sensors [9, 10], and antennas [11, 12]. Metasurface-induced extraordinary optical transmission (EOT) in the terahertz range is an active research field [13-16]. Bethe theory states that the transmission from a perforated metallic film is proportionate to $(r/\lambda)^4$ where $\lambda$ is the wavelength of the impinging electromagnetic field and $r$ is the radius of the hole. Contrary to Bethe's prediction, transmission resonances have been achieved through an array of subwavelength holes in a metallic film. This phenomenon is called EOT [17, 18].

Various methods have been employed to actively tune the extraordinary optical transmission. A tunable EOT at near 1550 nm has been achieved based on a metal-dielectric-graphene structure. By varying the Fermi level of the graphene layer, the modulation depth of transmission amplitude can reach 50% [19]. The amplitude of extraordinary transmission at about 0.44 THz through metallic ring apertures on a graphene layer can be modulated by about 50% [20]. Embedding graphene in a metallic ring-rod nested structure has been utilized to control the amplitude of EOTs at the frequencies of 1.77 and 2.42 THz [21]. The resonance frequency of the EOT has been tuned by using surface acoustic waves in a structure consisting of a periodical set of metallic stripes placed on a piezoelectric material. An acousto-optic variation of $\Delta n=0.02$ in the refractive index shifts the transmission resonance from 1461.5 to 1474.5 nm [22]. A transmission resonance through superconducting NbN subwavelength hole arrays has been demonstrated at the frequency of approximately 0.59 THz. The transmission level decreases from 0.98 to 0.48 when the operating temperature increases from 8.2 to 18 K [23]. One-dimensional gold gratings patterned on a 170 nm thick $VO_2$ have been demonstrated to modulate the EOT near wavelengths of 1550 nm. In the insulator phase of $VO_2$, the EOT is observed, however, the optical transmission extinguishes by up to 6 dB in the metallic phase. The electric current is applied to initiate the phase transition in the $VO_2$ film under the gratings via carrier injection and heating [24]. The modulation of the EOT has been presented based on a structure composed of a gold circular hole array on top of a phase-change material of $Ge_2Sb_2Te_5$. The state of the $Ge_2Sb_2Te_5$ is reversibly changed between the amorphous and crystalline state by a nanosecond pulsed laser leading to the 88% modulation of the transmission level at the frequency of 0.85 THz [25].

In this paper, we investigate a hybrid metasurface consisting of periodic square-loop-shaped holes in a metallic sheet connected where the holes are connected with strips of a phase-change material. Phase change materials exhibit large reversible changes in their optical properties and they have been exploited to design various reconfigurable metamaterials [26-29] and integrated photonic devices [30-32]. Here, we achieve tunable extraordinary optical transmission based on the phase transition of $VO_2$. The performance of the designed polarization-sensitive metasurface is evaluated numerically. For the normal incident *y*-polarization field, the resonance transmission redshifts from 0.88 to 0.81 THz upon insulator-to-metallic phase transition of $VO_2$. However, for the normal incident *x*-polarization field, the transmission resonance observed at 0.81 THz in the insulator phase effectively vanishes upon transition to the metallic phase. The multipole decomposition is utilized to determine the dominant multipole. We also examine the effect of geometrical parameters on the performance of the proposed structure. The designed metasurface can be employed as a tunable bandpass filter or a switch for *y*- or *x*-polarizations, respectively.

## 2. Designing polarization-sensitive EOT

The schematic of the designed metasurface composed of a gold film with an array of square-loop-shaped holes connected with $VO_2$ strips is shown in Fig. 1. The $SiO_2$ is chosen as the substrate with a thickness of $t_s$=75 µm. The thickness of the gold film and $VO_2$ strips is $t_m$=0.2 µm. The size of the square-shaped slits is determined with $L_1$=70 µm and $L_2$=100 µm while the period of the square unit cell is $P$=150 µm. The width of the $VO_2$ strips is $w$=10 µm. The Drude model of $\varepsilon_{VO_2}(\omega)=\varepsilon_\infty-\omega_p^2(\sigma)/(\omega^2+i\gamma\omega)$ is taken to describe the relative permittivity of $VO_2$ where $\varepsilon_\infty=12$ is the dielectric permittivity at the infinite frequencies. The plasma frequency at $\sigma$ is expressed as $\omega_p^2(\sigma)=\omega_p^2(\sigma_0)\sigma/\sigma_0$ where $\sigma_0=3\times10^5\Omega^{-1}cm^{-1}$ and $\omega_p(\sigma_0)=1.4\times10^{15} rad/s$ while the collision frequency is $\gamma=5.75\times10^{13} rad/s$ [33]. The $VO_2$ is in the insulator and metallic phases corresponding to $\sigma$ of $2\times10^2$ and $2\times10^5$ S/m at approximately 25°C and 85°C, respectively [29]. The Drude model of $\varepsilon_{Au}(\omega)=1-\omega_p^2/(\omega^2+i\gamma\omega)$ is also utilized to describe the optical properties of gold. The plasma frequency is $\omega_p=1.37\times10^{16} rad/s$ while the collision frequency is $\gamma=1.2\times10^{14} rad/s$ [34]. In the terahertz range, the $SiO_2$ is considered to be a lossless medium with a dielectric constant of 3.8 [34]. The designed metasurface is simulated with the full-wave frequency-domain solver of the CST Microwave Studio software. Unit cell conditions are applied in the *x*- and *y*-axes while open boundary conditions are applied in the *z*-axis. The normal *x*- and *y*-polarized plane waves are incident on the unit cell and the transmission is calculated as $|S_{21}|^2$ where $S_{21}$ is the scattering parameter. The transmission spectra of the metasurface for the insulator and metallic phases of $VO_2$ are illustrated in Fig. 2. For *y*-polarization and the insulator phase, the transmission resonance occurs at 0.88 THz with an amplitude of 0.75 and a narrow bandwidth of 0.07 THz. Upon phase transition of $VO_2$ to the metallic state, transmission peak redshifts to 0.81 THz with higher amplitude of 0.88 and broader bandwidth of 0.22 THz. Therefore, for *y*-polarization, the metasurface operates as a reconfigurable EOT device. However, for *x*-polarization, the EOT is observed in the insulator phase while the electromagnetic field is entirely reflected in the metallic phase. The EOT is located at the frequency of 0.81 THz with a peak of 0.88 and a bandwidth of 0.22 THz, in the insulator phase. In the metallic phase, the transmission level remains lower than 0.14 in the 0.5-1.1 THz range. The designed metasurface can switch on and off the EOT for *x*-polarization. We define the transmission bandwidth as a window with transmission levels higher than 0.5. The electric field distribution at the resonance frequencies for different polarizations of the incident field and different phases of $VO_2$ is displayed in Fig. 3.

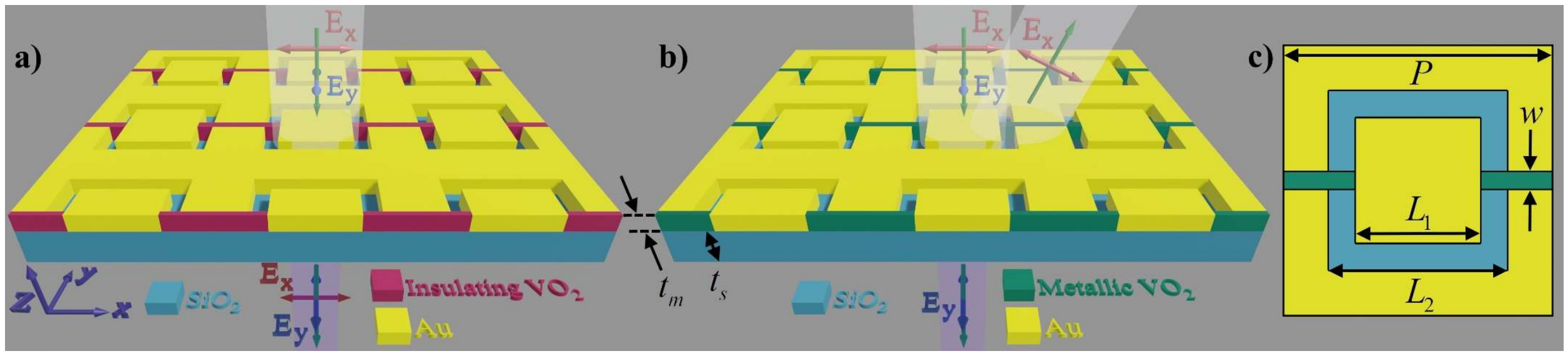

Fig. 1. Schematic of the proposed metasurface with $VO_2$ in a) the insulating phase and b) the metallic phase. c) Top view of the unit cell. The thickness of layers is not to scale. The direction of incident, reflected, and transmitted waves are shown with green arrows.

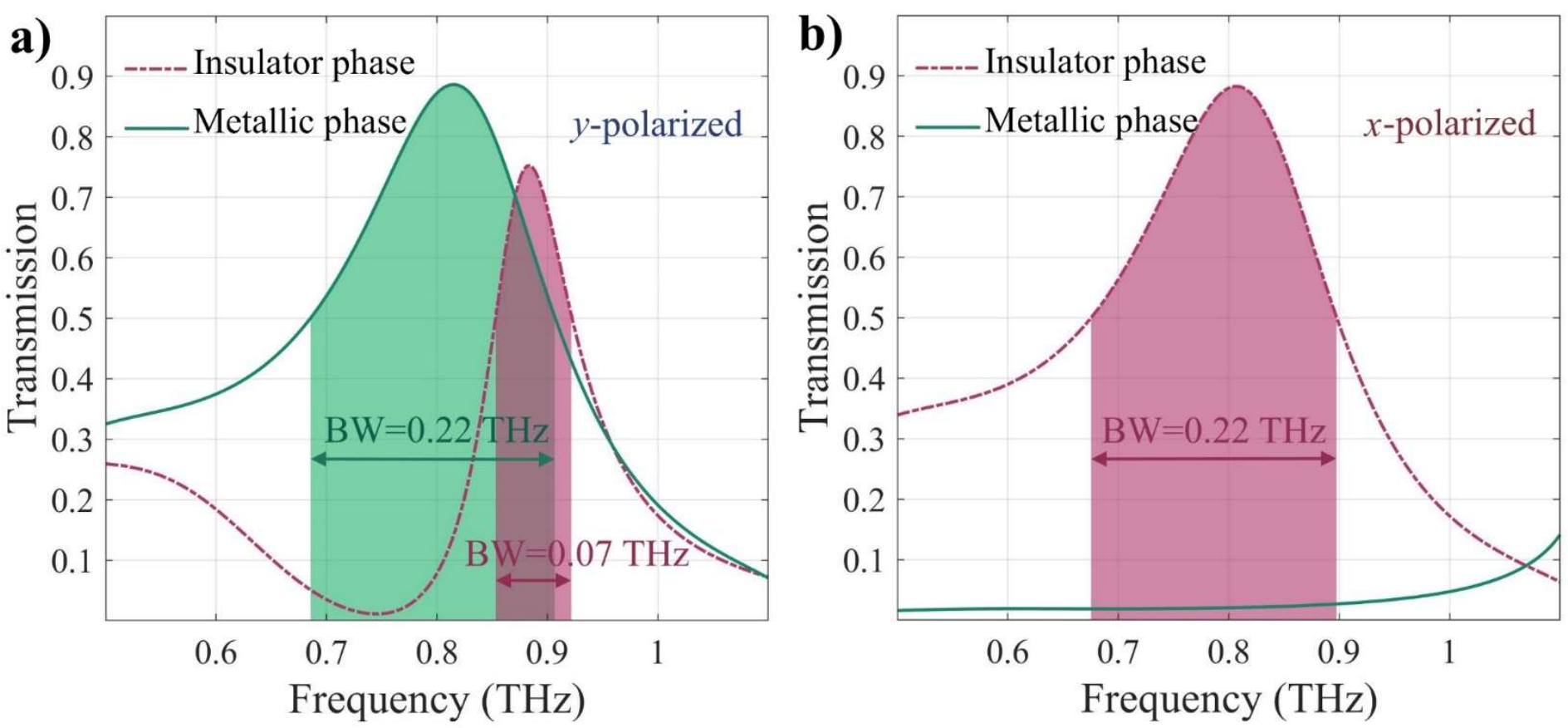

Fig. 2. The transmission spectra of the designed metasurface in the insulator and metallic phases of $VO_2$ for a) *y*-polarized and b) *x*-polarized incident terahertz field.

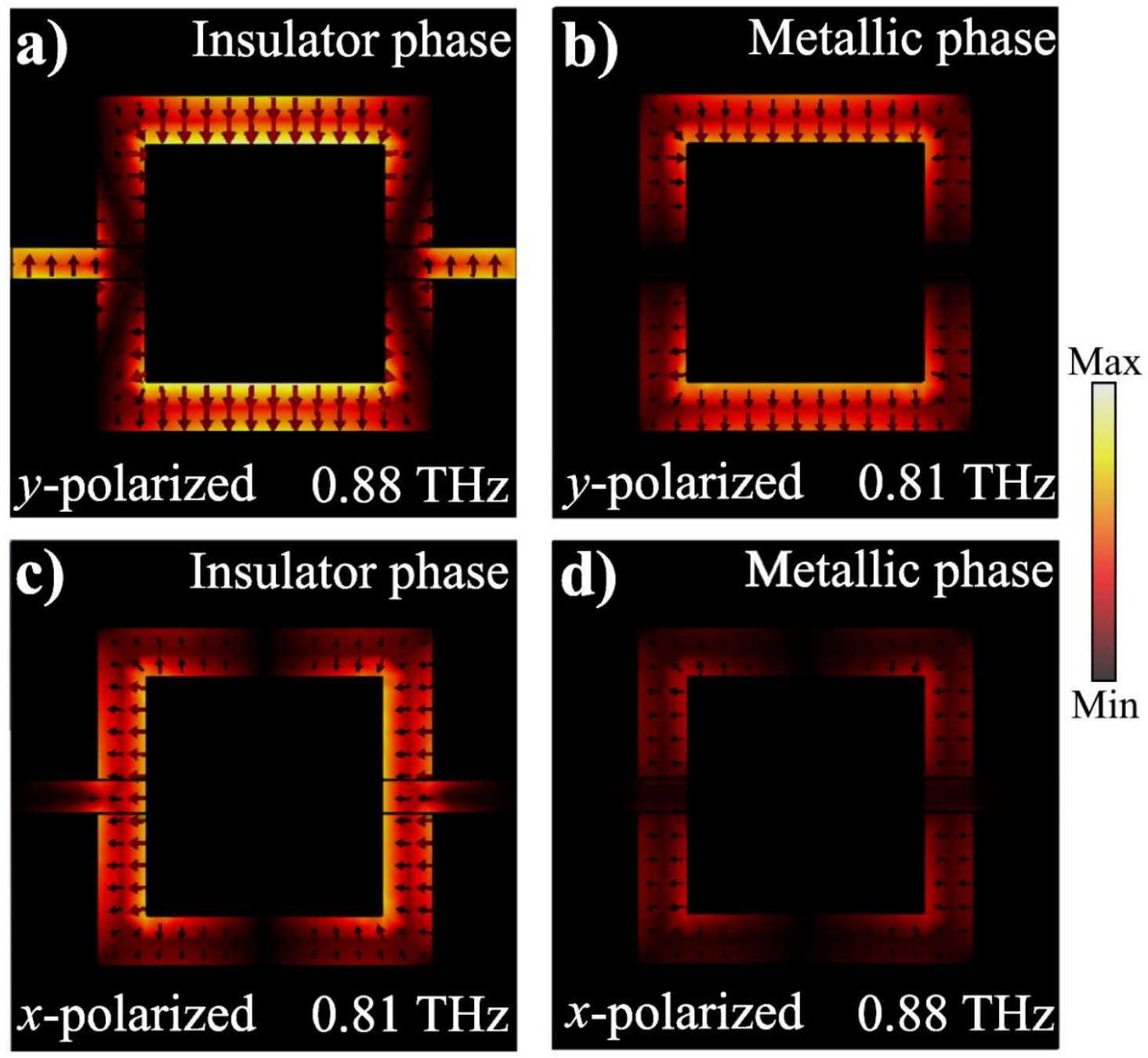

Fig. 3. The electric field distribution and vector at the resonance frequencies for different polarizations and $VO_2$ phases.

We also evaluate the scattering powers of multipoles on the transmission profile of the metasurface based on the Cartesian multipole decomposition. This method has been used to investigate the physical mechanism of resonances in metasurfaces [35, 36]. The multipole decomposition calculations are based on the induced current density, $\boldsymbol{J}(\boldsymbol{r})$ [37]

$$\boldsymbol{J}(\boldsymbol{r}) = -i\omega\varepsilon_0(n^2-1)\boldsymbol{E}(\boldsymbol{r}) \tag{1}$$

where $\boldsymbol{E}(\boldsymbol{r})$ is the electric field distribution and $n$ is the refractive index distribution while $\boldsymbol{r}$ is a position vector from the origin to point $(x, y, z)$. $\varepsilon_0$ is the permittivity of the free space and $\omega$ is the angular frequency. We calculate the electric dipole ($\boldsymbol{p}$), magnetic dipole ($\boldsymbol{m}$), electric quadrupole ($Q^e$), magnetic quadrupole ($Q^m$), and toroidal dipole ($\boldsymbol{T}$) [38]

$$\boldsymbol{p} = \frac{1}{i\omega}\int \boldsymbol{J} d^3\boldsymbol{r} \tag{2a}$$

$$\boldsymbol{m} = \frac{1}{2c}\int (\boldsymbol{r}\times\boldsymbol{J}) d^3\boldsymbol{r} \tag{2b}$$

$$\boldsymbol{Q}^{e}_{\alpha\beta} = \frac{1}{2i\omega}\int \left[ r_\alpha j_\beta + r_\beta j_\alpha - \frac{2}{3}(\boldsymbol{r}\cdot\boldsymbol{J})\delta_{\alpha,\beta} \right] d^3\boldsymbol{r} \tag{2c}$$

$$\boldsymbol{Q}^{m}_{\alpha\beta} = \frac{1}{2c}\int \left[ (\boldsymbol{r}\times\boldsymbol{J})_\alpha r_\beta + (\boldsymbol{r}\times\boldsymbol{J})_\beta r_\alpha \right] d^3\boldsymbol{r} \tag{2d}$$

$$\boldsymbol{T}_\alpha \approx \frac{1}{10c}\int \left[ (\boldsymbol{r}\cdot\boldsymbol{J}) r_\alpha - 2r^2 J_\alpha \right] d^3\boldsymbol{r} \tag{2e}$$

where $c$ is the speed of light and the summation indices ($\alpha$ and $\beta$) run over the Cartesian coordinates ($x$, $y$, and $z$). Here, the effect of other higher-order modes is negligible on the scattered intensity. The total scattering cross-section is [39, 40]

$$\begin{aligned} \sigma^{total}_{sca} &= \sigma^{p}_{sca} + \sigma^{m}_{sca} + \sigma^{Q^e}_{sca} + \sigma^{Q^m}_{sca} + \ldots \\ &= \frac{k^4}{6\pi\varepsilon_0^2 \left|E_{inc}\right|^2}\left[ \sum_\alpha \left( \left|p_\alpha + ikT_\alpha\right|^2 + \frac{\left|m_\alpha\right|^2}{c} \right) + \frac{1}{120}\sum_{\alpha\beta}\left( \left|kQ^{e}_{\alpha\beta}\right|^2 + \left|\frac{kQ^{m}_{\alpha\beta}}{c}\right|^2 \right) + \ldots \right] \end{aligned} \tag{3}$$

where $k$ is the wavenumber and $E_{inc}$ is the amplitude of the incident plane wave. We utilize the Lumerical FDTD to export the electric field and refractive index distribution. Afterward, the induced current density of Eq. 1 is calculated. The multipole moments of Eq. 2 are calculated by an open-source tool developed in MATLAB environment [37]. The contributions of each multipole moment to the scattering cross-section are shown in Figs. 4 and 5. The electric dipole is dominant and the effect of other multipoles on the performance of the metasurface is negligible. The electric field vectors in Fig. 3 show the dominant electric dipoles. Due to the differences such as calculation methods and simulation settings of CST and Lumerical software, the transmission spectra are slightly different. In the Lumerical simulations, for $y$-polarization, the resonance peaks are located at 0.88 and 0.74 THz in the insulator and metallic phases, respectively. For $x$-polarization and the insulator phase, the resonance peak occurs at 0.73 THz while no EOT is observed in the metallic phase. We also calculate the transmission of the metasurface based on the multipolar expansion using the generalized Kerker effect. The interference between the scattering fields from the multipoles in the forward direction can be used to calculate the transmission of the proposed metasurface. Considering the substrate effect on the transmission, the transmission coefficients for $x$- and $y$-polarizations are given by [41, 42]

$$t_x = [1 + \frac{ik_d}{2E_{inc}A\varepsilon_0\varepsilon_d}(p_x + \frac{1}{v_d}m_y - \frac{ik_d}{6}Q^{e}_{xz} - \frac{ik_d}{2v_d}Q^{m}_{yz})]t_0 e^{ik_d z_0} \tag{4a}$$

$$t_y = [1 + \frac{ik_d}{2E_{inc}A\varepsilon_0\varepsilon_d}(p_y - \frac{1}{v_d}m_x - \frac{ik_d}{6}Q^{e}_{yz} + \frac{ik_d}{2v_d}Q^{m}_{xz})]t_0 e^{ik_d z_0} \tag{4b}$$

where $A$ is the area of the unit cell and $z_0$ is the distance between the expansion center and substrate interface. Fresnel transmission coefficient of a normally incident field is $t_0 = 2\sqrt{\varepsilon_m}/(\sqrt{\varepsilon_m} + \sqrt{\varepsilon_s})$. $\varepsilon_s$, $\varepsilon_d$, and $\varepsilon_0$ are the permittivities of the substrate, surrounding medium, and vacuum, respectively. $k_d$ and $v_d$ are the wave number and speed of light in the surrounding medium, respectively. The transmission of the periodic metasurface is calculated by $T_{x,y} = \sqrt{\varepsilon_s / \varepsilon_d} \left| t_{x,y} \right|^2$. In Fig. 6, the transmission spectrum calculated by Lumerical as well as the transmission spectrum based on multipole scattering fields are shown for the *x*-polarization. Considering substrate-mediated inter-nanoantenna coupling, substrate-mediated self-coupling, and higher order multipoles may reduce the differences between the FDTD and multipole calculations [41].

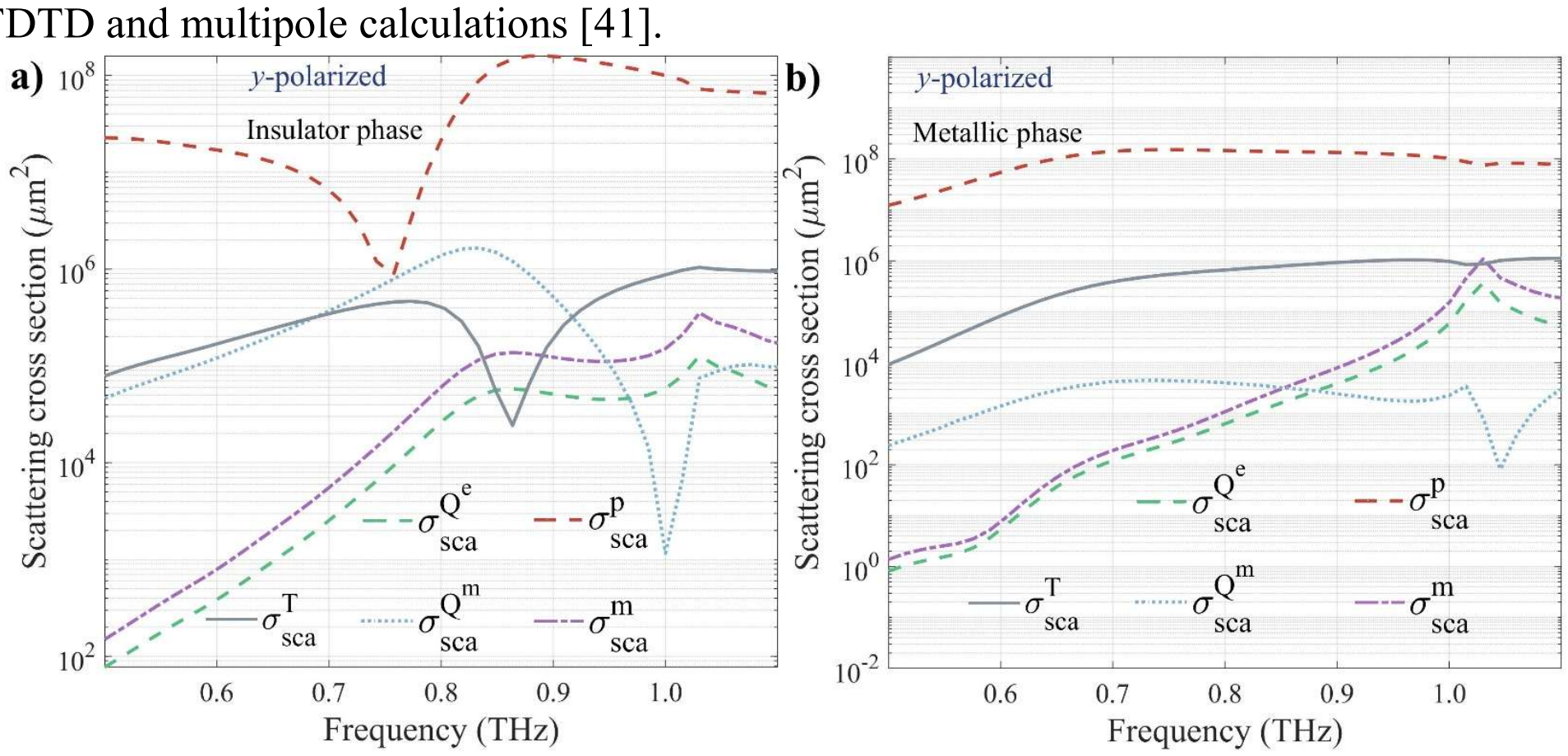

Fig. 4. Contribution of each multipole moment to the scattering cross-section for *y*-polarized incident field in a) the insulator and b) the metallic phases of $VO_2$.

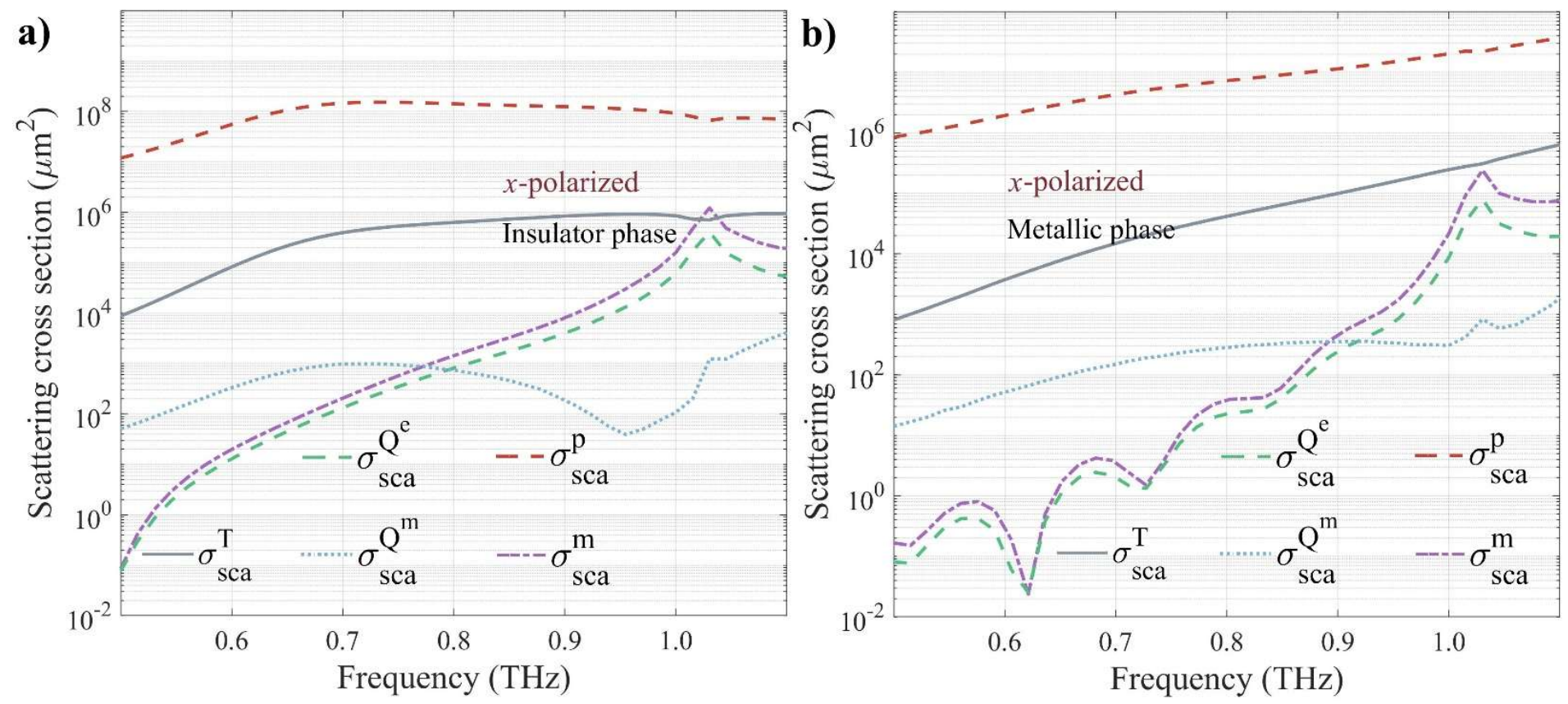

Fig. 5. Contribution of each multipole moment to the scattering cross-section for *x*-polarized incident field in a) the insulator and b) the metallic phases of $VO_2$.

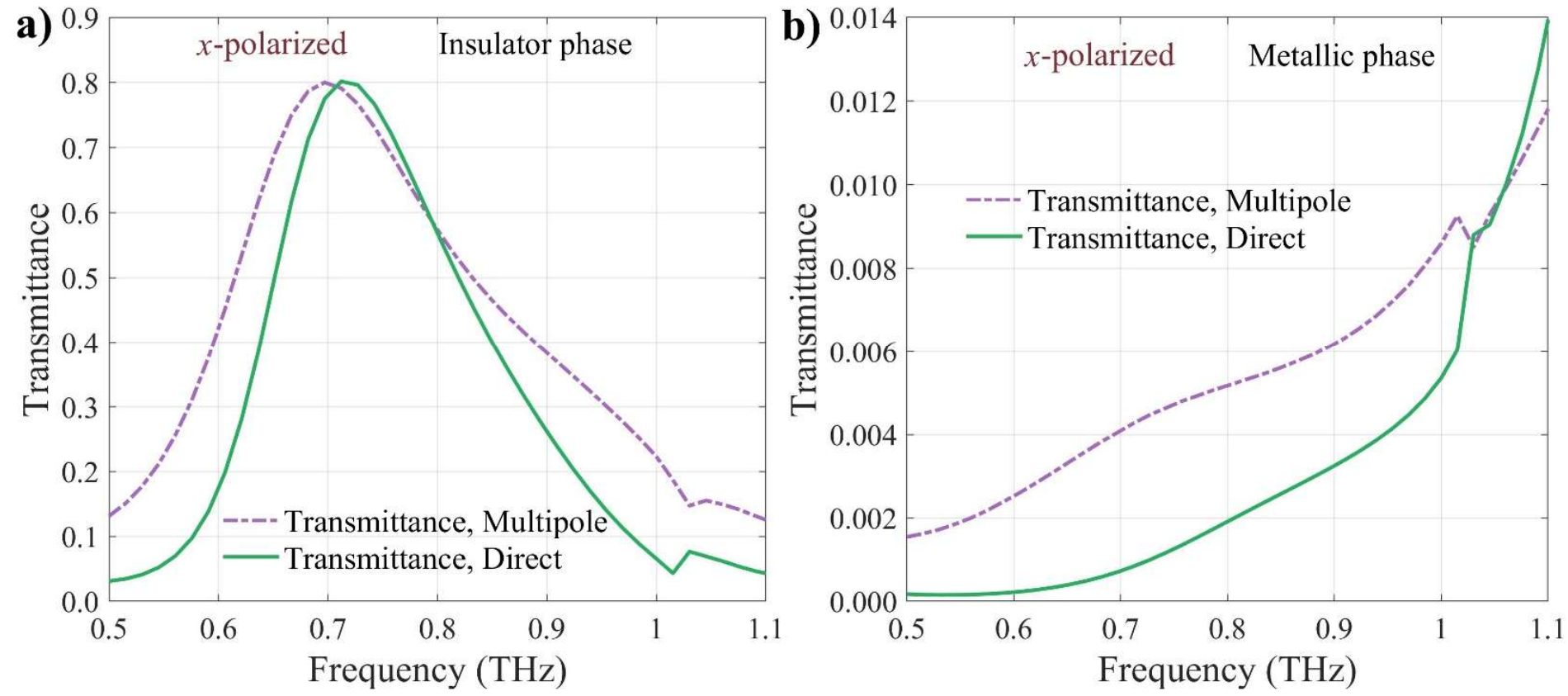

Fig. 6. Calculated transmission spectra through direct FDTD simulation compared with multipole interference induced transmission for *x*-polarized incident field in a) the insulator and b) the metallic phases of $VO_2$.

### *2.1 Effect of geometrical parameters*

The effect of the geometrical parameters on the resonant transmission is discussed in this subsection. First, the inner side length of the rectangular holes is varied while the other geometrical parameters are fixed to $t_s$=75 µm, $t_m$=0.2 µm, $L_2$=100 µm, $P$=150 µm, and $w$=10 µm. As shown in Fig. 7(a), for *y*-polarization, when the inner side length is set to $L_1$=60 µm, the transmission resonance peak is located at 0.94 and 0.89 THz with amplitudes of 0.89 and 0.92 in the insulator and metallic phases, respectively. The resonance frequency shifts by 0.05 THz upon phase transition of $VO_2$. For $L_1$=70 µm, the transmission resonance occurs at 0.88 and 0.81 THz while the peak amplitude is 0.75 and 0.88 in the insulator and metallic phases, respectively. In this case, the frequency shifts by 0.07 THz. Increasing the inner side length to $L_2$=80 µm blueshifts the resonances to 0.85 and 0.75 THz with amplitudes of 0.56 and 0.83 in the insulator and metallic phases, respectively. The frequency shift upon phase transition of $VO_2$ is 0.1 THz. As the $L_1$ increases, higher frequency shift can be achieved while the transmission amplitude decreases. As shown in Fig. 7(b), for *x*-polarization and in the insulator phase of $VO_2$, the transmission resonance blueshifts to 0.88, 0.81, and 0.74 THz while the peak amplitude decreases to 0.91, 0.88, and 0.82 as the inner side length increases to $L_1$=60, 70, and 80 µm, respectively. In the metallic phase, the transmission level increases to 0.08, 0.14, and 0.37 as the inner side length increases to $L_1$=60, 70, and 80 µm, respectively. Considering the trade-off between frequency shift and transmission amplitude, we chose $L_1$=70 µm.

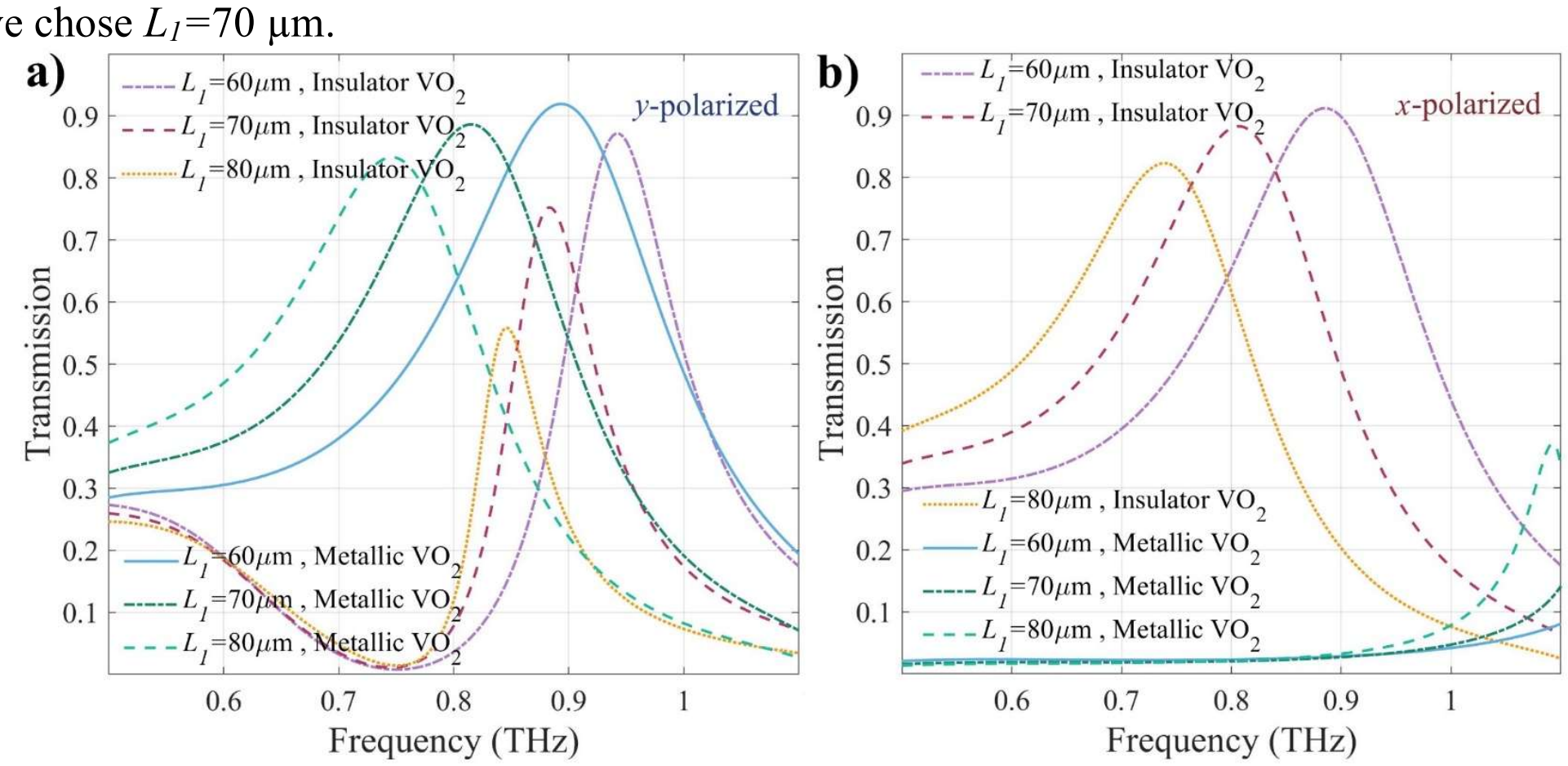

Fig. 7. Comparison of transmission spectra of the metasurface under different inner side lengths of square-shaped slits for a) *y*-polarized and b) *x*-polarized incident terahertz field.

Next, we examine the effect of $VO_2$ strip width on the EOT of the metasurface. As shown in Fig. 8(a), for *y*-polarization, increasing the strip width to *w*=5, 10, and 20 µm redshifts the transmission resonance to 0.87, 0.88, and 0.91 THz while the peak amplitude decreases to 0.76, 0.75, and 0.68 in the insulator phase, respectively. In the metallic phase of $VO_2$, the transmission resonance slightly redshifts to 0.81, 0.81, and 0.84 THz with amplitude of 0.93, 0.88, and 0.84, respectively. Upon insulator-to-metallic transition of $VO_2$, the frequency shift of 0.06, 0.07, and 0.07 THz is achieved for *w*=5, 10, and 20 µm, respectively. As shown in Fig. 8(b), for *x*-polarization, changing the strip width from *w*=5 to 20 µm has a limited effect on the frequency of the resonance peak while the transmission amplitude decreases from 0.92 to 0.84 in the insulator phase. In the metallic phase, the transmission level in the 0.5-1.1 THz remains below 0.31, 0.14, and 0.04 for *w*=5, 10, and 20 µm, respectively. We chose the strip width of *w*=10 µm since it offers relatively higher frequency shift and transmission amplitude. Other geometrical parameters also affect the EOT, however, we do not discuss them here for brevity.

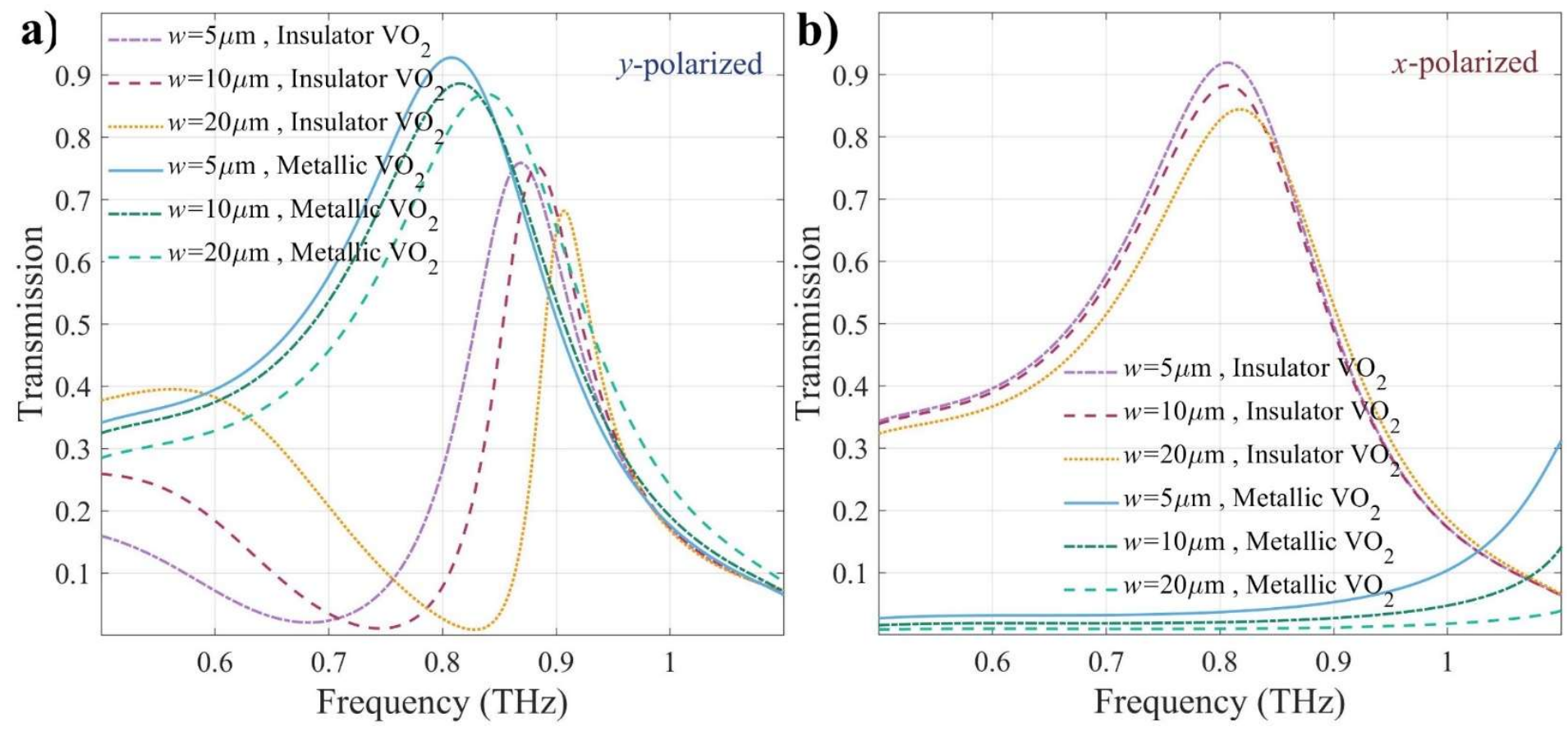

Fig. 8. Comparison of transmission spectra of the metasurface under different widths of $VO_2$ strips for a) *y*-polarized and b) *x*-polarized incident terahertz field.

We also examine the effect of unit cell's period on the EOT of the metasurface. We only change $P$ while other geometrical parameters are fixed to $t_s$=75 µm, $t_m$=0.2 µm, $L_2$=100 µm, $L_1$=70 µm, and $w$=10 µm. As shown in Fig. 9(a), for *y*-polarization, as the period increases from $P$=125 to 175 µm the transmission resonance redshifts from 0.94 to 0.84 THz in the insulator phase. In the metallic phase, the size of the hole remains constant, therefore, increasing the period results in lower transmission amplitude while the resonance frequency redshifts from 0.83 to 0.78 THz. Fig. 9(b) shows the transmission spectra for the x-polarization. Increasing the period causes the resonance frequency to redshift from about 0.82 THz to 0.77 THz in the insulator phase. In the metallic phase, the transmission level decreases from 0.04 to 0.01 at the frequency of 0.8 THz as the period of the unit cell increases from $P$=125 to 175 µm.

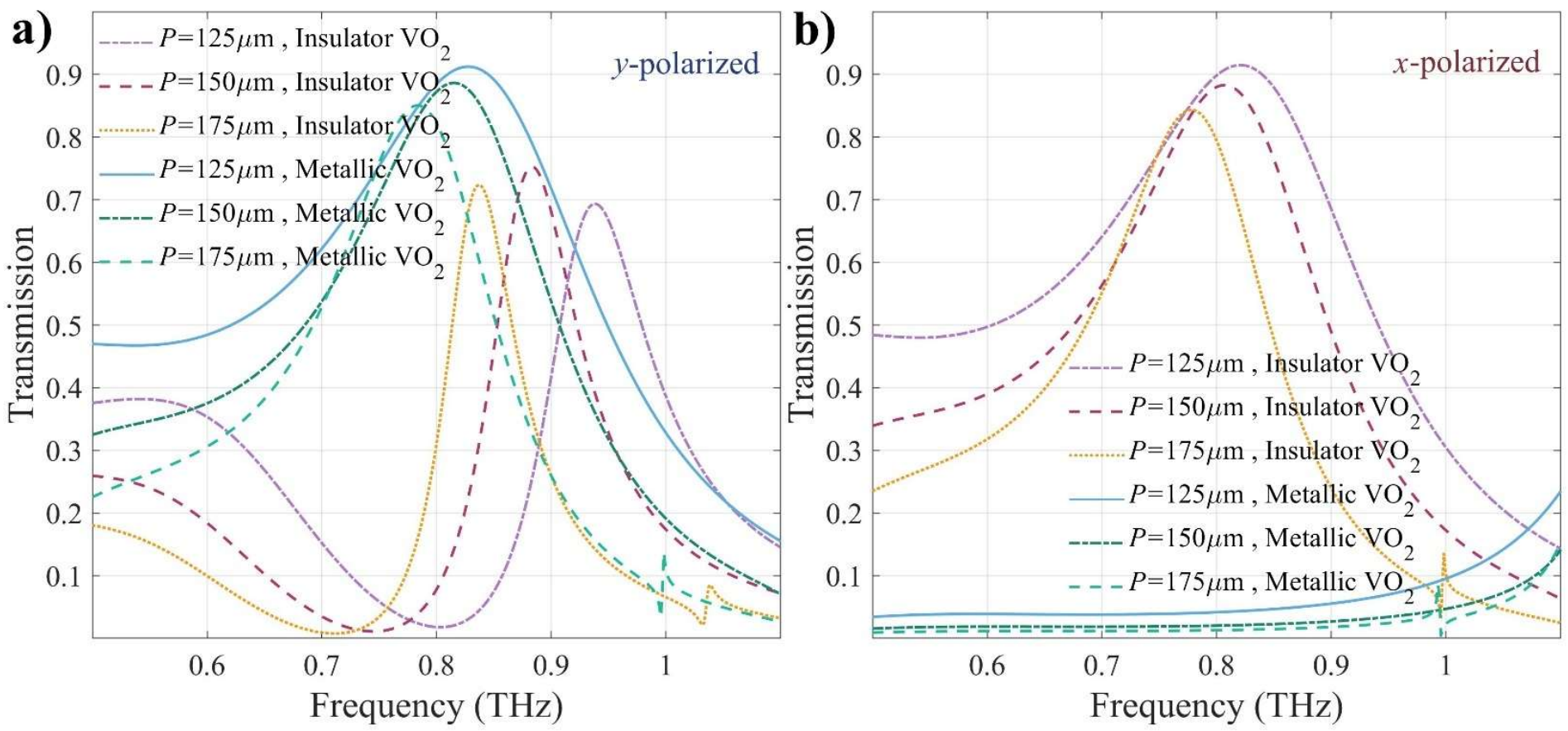

Fig. 9. Comparison of transmission spectra of the metasurface under different period of unit cell for a) *y*-polarized and b) *x*-polarized incident terahertz field.

## 3. Partial phase-transition of $VO_2$

At room temperature, the $VO_2$ strips are in the insulator phase. However, as the operating temperature of the metasurface increases the metallic domains of $VO_2$ form and expand gradually. In the intermediate states, where the insulator and metallic phases coexist, the effective conductivity of the $VO_2$ can be calculated by effective medium theory. Finally, at the temperature of about 85°C, the metallic domains coalesce into a continuous metallic film of $VO_2$ [43]. The effect of intermediate states during the phase transition of $VO_2$ on the performance of the EOT is illustrated in Fig. 10. For *y*-polarization, as the conductivity of the $VO_2$ increases, the transmission peak slightly blueshifts while its amplitude decreases initially. Then, the transmission peak redshifts and its amplitude increases. For $2\times10^2$, $6\times10^2$, and $1\times10^3$ S/m, the transmission peak occurs at 0.88, 0.90, and 0.91 THz with an amplitude of 0.75, 0.52, and 0.41, respectively. However, as the conductivity increases to $2\times10^3$, $3\times10^3$, $5\times10^3$, $1\times10^4$, $5\times10^4$, and $2\times10^5$

S/m, the transmission peak slightly redshifts to 0.86, 0.82, 0.81, 0.81, 0.81, and 0.81 THz while its amplitude increases to 0.39, 0.50, 0.63, 0.76, 0.86, and 0.88, respectively. As shown in Fig. 10(b), for $2\times10^2$, $6\times10^2$, $1\times10^3$, $2\times10^3$, $3\times10^3$, $5\times10^3$, and $1\times10^4$ S/m, the transmission resonance gradually blueshifts to 0.81, 0.81, 0.81, 0.82, 0.83, 0.86, and 0.95 THz with lower amplitude of 0.88, 0.77, 0.68, 0.52, 0.41, 0.29, and 0.17, respectively. For conductivities of $5\times10^4$ and $2\times10^5$, the transmission peak occurs outside of the studied range of 0.5-1.1 THz. The modulation depth of the metasurface is calculated as $(A_1-A_2)\times100/(A_1+A_2)$, where $A_1$ and $A_2$ are the transmission peaks at the resonance frequency before and after phase transition of $VO_2$. For the *x*-polarization, the modulation depth of about 68% can be achieved.

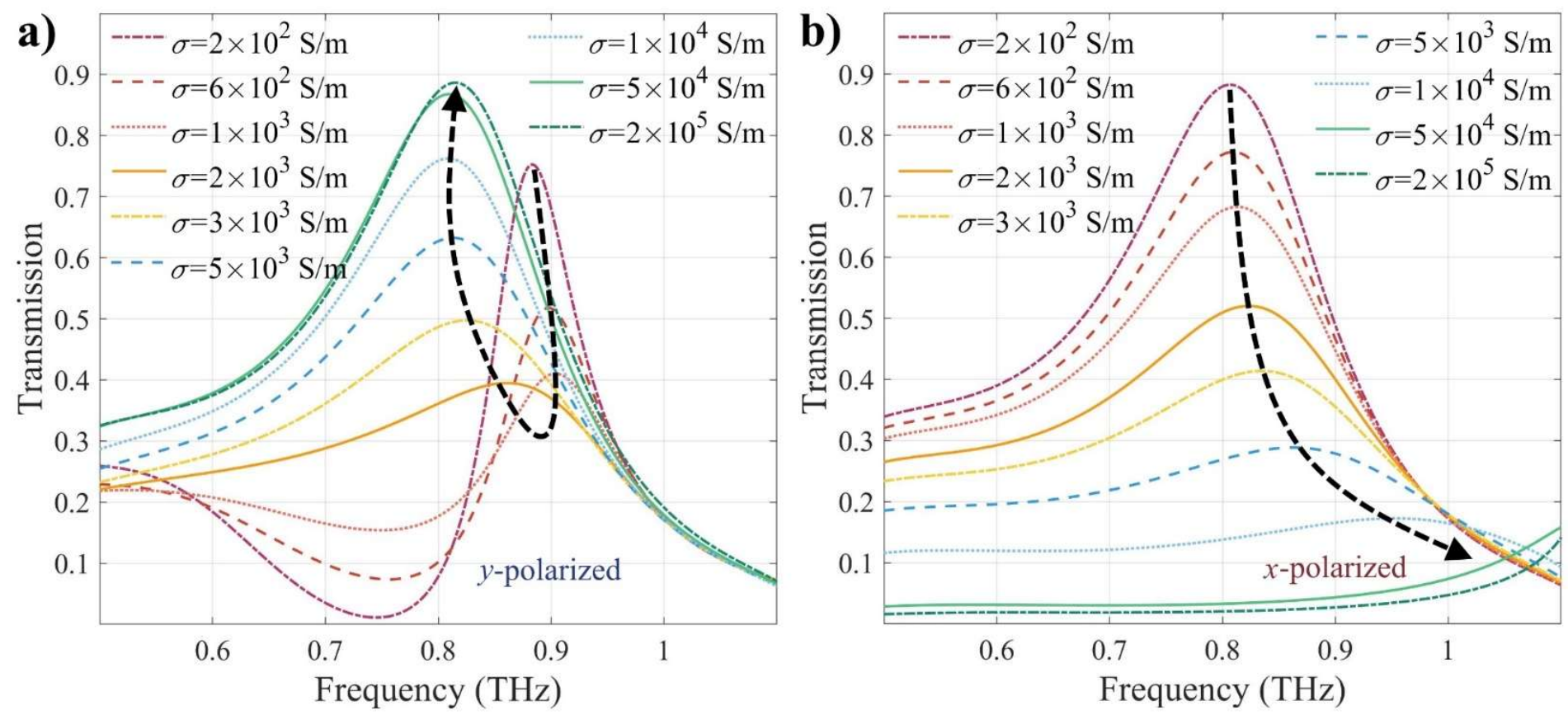

Fig. 10. The transmission spectra of the metasurface with different conductivities of $VO_2$ for a) *y*-polarized and b) *x*-polarized incident terahertz field.

## 4. Angle dependence of EOT

Low sensitivity to incident angle of metasurfaces is crucial. Hence, in this subsection, we investigate the performance of the designed metasurface by varying the incident angle from 0° to 85° in 5° steps. Fig. 11 shows the transmission spectra of the metasurface for the *y*-polarized wave and different phases of $VO_2$. In the insulator state of $VO_2$, as the incident angle increases from 0° to 40°, the transmission amplitude increases from about 0.71 to 0.78 while the resonance frequency shifts from 0.88 to 0.81 THz. Further increase in the incident angle results in deceasing the transmission amplitude to 0.44 while the resonance frequency remains at about 0.78 THz. At incident angles larger than 25°, a second transmission peak appears in the studied frequency range. This transmission peak shifts from 1.08 to 0.92 THz with a decrease in amplitude from 0.66 to 0.34. In the metallic phase, the transmission resonance shifts from 0.81 to 0.79 THz as the incident angle increases from 0° to 85°. Meanwhile, the transmission amplitude drops from about 0.90 to 0.37 while the full-width at half-maximum (FWHM) decreases from 0.26 to 0.02 THz. Similar to the insulator phase, a second transmission peak appears at about 1.08 THz with the amplitude of 0.79 which shifts to 0.92 THz with the amplitude of 0.43.

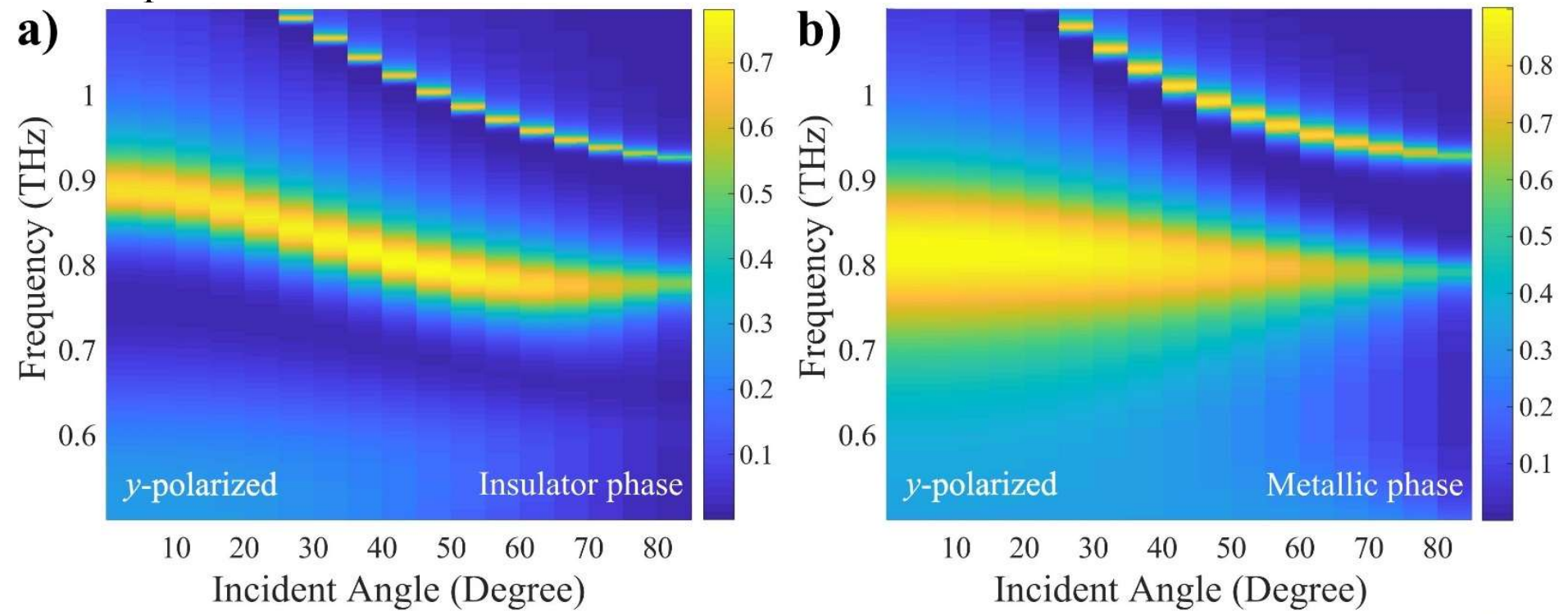

Fig. 11. Transmission spectra versus the incident angle for the *y*-polarized field in a) the insulator and b) the metallic phases of $VO_2$.

## 5. Conclusion

In summary, we propose a hybrid metasurface that exhibits polarization-sensitive extraordinary optical transmission. The metasurface is composed of a periodic array of square apertures in a gold film while the holes are connected with $VO_2$ strips. When the *y*-polarized terahertz field is incident on the metasurface, upon insulator-to-metal transition of $VO_2$, the transmission resonance redshifts from 0.88 to 0.81 THz. However, for *x*-polarization, the EOT is efficiently suppressed in the metallic phase while the EOT can be observed at the frequency of 0.81 THz in the insulator phase. The designed metasurface can be employed as a reconfigurable filter, modulator, or switch.

**Disclosures.** The authors declare no conflicts of interest.

**Data availability.** Data underlying the results presented in this paper are not publicly available at this time but may be obtained from the authors upon reasonable request.